\documentclass[fleqn,twocolumn,showpacs,preprintnumbers,showkeys]{revtex4}
\usepackage{graphicx}
\usepackage{amssymb}
\usepackage{amsmath}
\usepackage{bm}
\usepackage[flushleft]{caption2}

\begin{document}
\setcounter{page}{1}
\title{Field equations of electromagnetic and gravitational fields}
\author{Zihua Weng}
\email{xmuwzh@xmu.edu.cn.}
\affiliation{School of Physics and
Mechanical \& Electrical Engineering,
\\Xiamen University, Xiamen 361005, China}

\begin{abstract}
The paper studies the validity of Maxwell equation in the case for
coexistence of electromagnetic field and gravitational field. With
the algebra of quaternions, the Newton's law of gravitation is the
same as that in classical theory of gravitational field. Meanwhile
the Maxwell equation is identical with that in classical theory of
electromagnetic field. And the related conclusions can be spread to
the case for coexistence of electromagnetic field and gravitational
field by the algebra of octonions. The study claims that Maxwell
equation keeps unchanged in the case for coexistence of
gravitational field and electromagnetic field, except for the
direction of displacement current.
\end{abstract}



\maketitle

\section{INTRODUCTION}

The validity of Maxwell equation is being doubted all the time in
the gravitational field and electromagnetic field. And this validity
remains as puzzling as ever. The existing theories do not explain
why Maxwell equation keeps unchanged, and then do not offer
compelling reason for this special situation. The paper attempts to
find out why Maxwell equation keeps the same in most cases, even in
the gravitational field.

The electromagnetic field can be described with the quaternion,
which was invented by W. R. Hamilton \cite{hamilton, adler} in 1843.
The quaternion was first used by J. C. Maxwell to represent field
equations and properties of electromagnetic field in 1861
\cite{maxwell, dirac}. O. Heaviside in 1884 recast Maxwell equation
in terms of vector terminology and electromagnetic forces
\cite{jackson}, thereby reduced the original twenty equations down
to the four differential equations in classical electromagnetic
field \cite{gogberashvili, dzhunushaliev}.

Recently, the algebra of quaternions can be used to represent the
gravitational field \cite{weng}. In 1687, I. Newton published the
mechanical theory to describe the three laws of motion and the
universal gravitation \cite{newton}. In 1812, S.-D. Poisson
reformulated Newton's law of gravitation in terms of the scalar
potential \cite{weinberg}.

In the paper, by means of the feature of quaternions, we obtain
Maxwell equation in the electromagnetic field, and Newton's law of
gravitation in the gravitational field. Further, making use of the
algebra of octonions, we can achieve consistently Maxwell equation
and some other equations in the case for coexistence of
gravitational field and electromagnetic field.

\section{Gravitational field}

The feature of gravitational field can be described by the algebra
of quaternions. In the quaternion space, the coordinates are $r_0$,
$r_1$, $r_2$, and $r_3$, with the basis vector $\mathbb{E}_g = (1 ,
\emph{\textbf{i}}_1 , \emph{\textbf{i}}_2 , \emph{\textbf{i}}_3)$.
The radius vector $\mathbb{R}_g = \Sigma (r_i \emph{\textbf{i}}_i)$,
and the velocity $\mathbb{V}_g = \Sigma (v_i \emph{\textbf{i}}_i)$,
with $\emph{\textbf{i}}_0 =1 $ . Where, $r_0 = v_0 t$; $v_0$ is the
speed of gravitational intermediate boson, and $t$ is the time. $j =
1, 2, 3$ ; $i = 0, 1, 2, 3 $ .

The gravitational potential is,
\begin{eqnarray}
\mathbb{A}_g = \Sigma (a_i \emph{\textbf{i}}_i)
\end{eqnarray}
and the strength $\mathbb{B}_g = \Sigma (b_i \emph{\textbf{i}}_i)$
of gravitational field is
\begin{eqnarray}
\mathbb{B}_g = \lozenge \circ \mathbb{A}_g~.
\end{eqnarray}
where, the $\circ$ denotes the quaternion multiplication; the
operator $ \lozenge = \Sigma (\emph{\textbf{i}}_i \partial_i) ;
~\partial_i =
\partial/\partial r_i$ .

The gravitational strength $\mathbb{B}_g$ covers two components,
$\textbf{g}/v_0 = \partial_0 \textbf{a} + \nabla a_0 $ and
$\textbf{b} = \nabla \times \textbf{a} $ .
\begin{eqnarray}
\textbf{g}/v_0 = && \emph{\textbf{i}}_1 ( \partial_0 a_1 +
\partial_1 a_0 ) + \emph{\textbf{i}}_2 ( \partial_0 a_2 + \partial_2
a_0 )
\nonumber\\
&& + \emph{\textbf{i}}_3 ( \partial_0 a_3 + \partial_3 a_0 )
\\
\textbf{b} = && \emph{\textbf{i}}_1 ( \partial_2 a_3 -
\partial_3 a_2 ) + \emph{\textbf{i}}_2 ( \partial_3 a_1 - \partial_1
a_3 )
\nonumber\\
&& + \emph{\textbf{i}}_3 ( \partial_1 a_2 - \partial_2 a_1 )
\end{eqnarray}
where, the gauge equation $b_0 = \partial_0 a_0 + \nabla \cdot
\textbf{a} = 0$; the vectorial potential $\textbf{a} = \Sigma (a_j
\emph{\textbf{i}}_j)$; $\nabla = \Sigma (\emph{\textbf{i}}_j
\partial_j)$.

\begin{table}[b]
\caption{\label{tab:table1}The quaternion multiplication table.}
\begin{ruledtabular}
\begin{tabular}{ccccc}
& 1 & $\emph{\textbf{i}}_1$ & $\emph{\textbf{i}}_2$ &
$\emph{\textbf{i}}_3$  \\ \colrule 1 & 1 & $\emph{\textbf{i}}_1$ &
$\emph{\textbf{i}}_2$ &
$\emph{\textbf{i}}_3$  \\
$\emph{\textbf{i}}_1$ & $\emph{\textbf{i}}_1$ & -1 &
$\emph{\textbf{i}}_3$  & $-\emph{\textbf{i}}_2$ \\
$\emph{\textbf{i}}_2$ & $\emph{\textbf{i}}_2$ &
$-\emph{\textbf{i}}_3$ & -1 & $\emph{\textbf{i}}_1$ \\
$\emph{\textbf{i}}_3$ & $\emph{\textbf{i}}_3$ &
$\emph{\textbf{i}}_2$ & $-\emph{\textbf{i}}_1$ & -1 \\
\end{tabular}
\end{ruledtabular}
\end{table}

The linear momentum density, $\mathbb{P}= m \mathbb{V}_g$, is the
source density $\mathbb{S}_g$ of gravitational field. The latter one
is defined from the gravitational strength $\mathbb{B}_g$.
\begin{eqnarray}
\lozenge^* \circ \mathbb{B}_g = - \mu_g \mathbb{S}_g
\end{eqnarray}
where, $m$ is the mass density; $*$ denotes the quaternion
conjugate; $\mu_g = 4 \pi G / v_0^2$ is one coefficient, and $G$ is
the gravitational constant.

The strength $\textbf{b}$ is too weak to detect. For example, the
mass $m_1$, $m_2$ move with the speed $v_1$, $v_2$ respectively.
They move parallel, and there will be two forces $f_1$, $f_2$
between them. Wherein, $f_1 = m_2 g_1$, with $g_1 = G m_1 / r^2$;
and $f_2 = m_2 b_1 v_2 $, with $b_1 < (\mu_g/4\pi) m_1 v_1 / r^2$.
In most cases, $f_2/f_1 \approx v_1 v_2 / c^2 \ll 1$, therefore we
will neglect $f_2$ generally. In the Newtonian gravity theory, there
are $\textbf{a} = 0$, $\textbf{b} = 0$, and $f_2 = 0$ specially.

\subsection{Newton's law of gravitation}

In the gravitational field, the scalar part $s_0$ of source
$\mathbb{S}_g$ in Eq.(5) is rewritten as follows.
\begin{eqnarray}
\nabla^* \cdot \textbf{h} = - \mu_g s_0
\end{eqnarray}
where, $\textbf{h} = \Sigma (b_j \emph{\textbf{i}}_j)$; $s_0 = p_0 =
m v_0$.

Further, the above is reduced to
\begin{eqnarray}
\nabla^* \cdot ( \textbf{g} / v_0 + \textbf{b} ) = - \mu_g m v_0~.
\end{eqnarray}

Eqs.(2) and (4) yield the equation,
\begin{eqnarray}
\nabla \cdot \textbf{b} = 0
\end{eqnarray}
and then, we have Newton's law of gravitation.
\begin{eqnarray}
\nabla^* \cdot \textbf{g} = - m / \varepsilon_g
\end{eqnarray}
where, the coefficient $\varepsilon_g = 1/(\mu_g v_0^2)$.

The above states that the gravitational potential $\textbf{a}$ has
an influence on the Newton's law of gravitation, although the term
$\nabla \cdot \partial_0 \textbf{a}$ is very tiny. Meanwhile
Newton's law of gravitation is an invariant under Galilean
transformation and Lorentz transformation, and the definition of
gravitational strength can be extended from the steady state to
movement state.

\begin{table}[b]
\caption{\label{tab:table1}The operator and multiplication of the
physical quantity in the quaternion space.}
\begin{ruledtabular}
\begin{tabular}{ll}
definition                  &   meaning                                              \\
\hline
$\nabla \cdot \textbf{a}$   &  $-(\partial_1 a_1 + \partial_2 a_2 + \partial_3 a_3)$ \\
$\nabla \times \textbf{a}$  &  $\emph{\textbf{i}}_1 ( \partial_2 a_3
                                 - \partial_3 a_2 ) + \emph{\textbf{i}}_2 ( \partial_3 a_1
                                 - \partial_1 a_3 )$                                 \\
$$                          &  $ + \emph{\textbf{i}}_3 ( \partial_1 a_2
                                 - \partial_2 a_1 )$                                 \\
$\nabla a_0$                &  $\emph{\textbf{i}}_1 \partial_1 a_0
                                 + \emph{\textbf{i}}_2 \partial_2 a_0
                                 + \emph{\textbf{i}}_3 \partial_3 a_0  $             \\
$\partial_0 \textbf{a}$     &  $\emph{\textbf{i}}_1 \partial_0 a_1
                                 + \emph{\textbf{i}}_2 \partial_0 a_2
                                 + \emph{\textbf{i}}_3 \partial_0 a_3  $             \\
\end{tabular}
\end{ruledtabular}
\end{table}

\subsection{Ampere's law of gravitation}

In the quaternion space, the vectorial part $\textbf{s}$ of linear
momentum density $\mathbb{S}_g$ can be decomposed from Eq.(5).
\begin{eqnarray}
\partial_0 \textbf{h} + \nabla^* \times \textbf{h} = - \mu_g
\textbf{s}
\end{eqnarray}
where, $\textbf{s} = \Sigma (s_j \emph{\textbf{i}}_j) $; $s_j = p_j
= m v_j $.

The above can be rewritten as follows.
\begin{eqnarray}
\partial_0 ( \textbf{g} / v_0 + \textbf{b} ) + \nabla^* \times ( \textbf{g} / v_0 + \textbf{b} )
= - \mu_g \textbf{s}
\end{eqnarray}

Eqs.(2) $-$ (4) yield the equation,
\begin{eqnarray}
\partial_0 \textbf{b} + \nabla^* \times \textbf{g} / v_0 = 0
\end{eqnarray}
and then, we obtain Ampere's law of gravitation.
\begin{eqnarray}
\partial_0 \textbf{g} / v_0 + \nabla^* \times \textbf{b} = - \mu_g \textbf{s}
\end{eqnarray}

The above means the Newton's law of gravitation in the quaternion
space is the same as that in classical gravitational theory. In the
quaternion space, the masses in either steady state or movement
state can exert the gravity on other objects. The linear momentum
yields the gravitational strength $\textbf{b}$, which may be quite
weak. And the strength $\textbf{b}$ and $\textbf{g}$ can be induced
each other in the gravitational field from Eq.(12).

\section{Electromagnetic field}

The feature of electromagnetic field can be represented by the
algebra of quaternions as well. In the quaternion space, the
coordinates are $r_0$, $r_1$, $r_2$, and $r_3$, with the basis
vector $\mathbb{E}_q = (1 , \emph{\textbf{i}}_1 ,
\emph{\textbf{i}}_2 , \emph{\textbf{i}}_3)$. The radius vector
$\mathbb{R}_q = \Sigma (r_i \emph{\textbf{i}}_i)$, and the velocity
$\mathbb{V}_q = \Sigma (v_i \emph{\textbf{i}}_i)$ . Where, $r_0 =
v_0 t$; $v_0$ is the speed of electromagnetic intermediate boson,
and $t$ is the time.

The electromagnetic potential is,
\begin{eqnarray}
\mathbb{A}_q = \Sigma (A_i \emph{\textbf{i}}_i)
\end{eqnarray}
and the electromagnetic strength $\mathbb{B}_q = \Sigma (B_i
\emph{\textbf{i}}_i)$ is
\begin{eqnarray}
\mathbb{B}_q = \lozenge \circ \mathbb{A}_q~.
\end{eqnarray}
where, $\textbf{A} = \Sigma (A_j \emph{\textbf{i}}_j)$.

The electromagnetic strength $\mathbb{B}_q$ includes two parts,
$\textbf{E}_q/v_0 = \partial_0 \textbf{A} + \nabla A_0 $ and
$\textbf{B}_q = \nabla \times \textbf{A} $ .
\begin{eqnarray}
\textbf{E}_q/v_0 = && \emph{\textbf{i}}_1 ( \partial_0 A_1 +
\partial_1 A_0 ) + \emph{\textbf{i}}_2 ( \partial_0 A_2 + \partial_2
A_0 )
\nonumber\\
&& + \emph{\textbf{i}}_3 ( \partial_0 A_3 + \partial_3 A_0 )
\\
\textbf{B}_q = && \emph{\textbf{i}}_1 ( \partial_2 A_3 -
\partial_3 A_2 ) + \emph{\textbf{i}}_2 ( \partial_3 A_1 - \partial_1
A_3 )
\nonumber\\
&& + \emph{\textbf{i}}_3 ( \partial_1 A_2 - \partial_2 A_1 )
\end{eqnarray}
where, the gauge equation $B_0 = \partial_0 A_0 + \nabla \cdot
\textbf{A} = 0$.

The electric current density $\mathbb{S}_q = q \mathbb{V}_q$ is the
source density of electromagnetic field, and is defined from the
electromagnetic strength $\mathbb{B}_q$ .
\begin{eqnarray}
\lozenge^* \circ \mathbb{B}_q = - \mu_q \mathbb{S}_q
\end{eqnarray}
where, $q$ is the electric charge density; $\mu_q$ is the
electromagnetic constant.

The above equation is same as that in classical theory of
electromagnetic field.

\subsection{Gauss's law}

In the electromagnetic field, the scalar part $S'_0$ of the source
$\mathbb{S}_q$ in Eq.(18) is rewritten as follows.
\begin{eqnarray}
\nabla^* \cdot \textbf{H}_q = - \mu_q S'_0
\end{eqnarray}
where, $\textbf{H}_q = \Sigma (B_j \emph{\textbf{i}}_j)$; $S'_0 = q
v_0$.

Further, the above is reduced to
\begin{eqnarray}
\nabla^* \cdot ( \textbf{E}_q / v_0 + \textbf{B}_q ) = - \mu_q q
v_0~.
\end{eqnarray}

Eqs.(15) and (17) yield the Gauss's law for magnetism,
\begin{eqnarray}
\nabla \cdot \textbf{B}_q = 0
\end{eqnarray}
and then, we have Gauss's law as follows.
\begin{eqnarray}
\nabla^* \cdot \textbf{E}_q = - q / \varepsilon_q
\end{eqnarray}
where, the coefficient $\varepsilon_q = 1/(\mu_q v_0^2)$.

By comparison with the Maxwell equation, we find that Eqs.(21) and
(22) are the same as that in Maxwell equation of classical
electromagnetic theory, although the definition of the gauge
equation $B_0 = 0$ is different.

\subsection{Ampere-Maxwell law}

In the quaternion space, the vectorial part $\textbf{S}_q$ of
electric current density $\mathbb{S}_q$ can be decomposed from
Eq.(18).
\begin{eqnarray}
\partial_0 \textbf{H}_q + \nabla^* \times \textbf{H}_q = - \mu_q
\textbf{S}_q
\end{eqnarray}
where, $\textbf{S}_q = \Sigma (S'_j \emph{\textbf{i}}_j) $; $S'_j =
q v_j $.

The above can be rewritten as follows.
\begin{eqnarray}
&& \partial_0 ( \textbf{E}_q / v_0 + \textbf{B}_q )
\nonumber\\
&& + \nabla^* \times ( \textbf{E}_q / v_0 + \textbf{B}_q ) = - \mu_q
\textbf{S}_q
\end{eqnarray}

Eqs.(15) $-$ (17) yield Faraday's law of induction,
\begin{eqnarray}
\partial_0 \textbf{B}_q + \nabla^* \times \textbf{E}_q / v_0 = 0
\end{eqnarray}
and then, we obtain Ampere-Maxwell law in the electromagnetic field
as follows.
\begin{eqnarray}
\partial_0 \textbf{E}_q / v_0 + \nabla^* \times \textbf{B}_q = -
\mu_q \textbf{S}_q
\end{eqnarray}

The above means that Eqs.(21) and (22) are combined with Eqs.(25)
and (26) to become Maxwell equation. By contrast with Maxwell
equation, we find that Eqs.(25) and (26) are the same as that in
classical electromagnetic theory, except for the direction of
displacement current.

\section{Electromagnetic field and gravitational field}

The feature of gravitational field and electromagnetic field can be
described simultaneously by the octonion space, which is consist of
two quaternion spaces.

In the quaternion space for the gravitational field, the basis
vector is $\mathbb{E}_g$, the radius vector is $\mathbb{R}_g$, and
the velocity is $\mathbb{V}_g$. In the quaternion space for the
electromagnetic field, the basis vector is $\mathbb{E}_e$ =
($\emph{\textbf{I}}_0$, $\emph{\textbf{I}}_1$,
$\emph{\textbf{I}}_2$, $\emph{\textbf{I}}_3$), the radius vector is
$\mathbb{R}_e = \Sigma (R_i \emph{\textbf{I}}_i)$, and the velocity
is $\mathbb{V}_e = \Sigma (V_i \emph{\textbf{I}}_i)$. The
$\mathbb{E}_e$ is independent of the $\mathbb{E}_g$, with
$\mathbb{E}_e$ = $\mathbb{E}_g$ $\circ$ $\emph{\textbf{I}}_0$ .

These two quaternion spaces can be combined together to become one
octonion space \cite{cayley, baez}, with the octonion basis vector
$\mathbb{E} = (1, \emph{\textbf{i}}_1, \emph{\textbf{i}}_2,
\emph{\textbf{i}}_3, \emph{\textbf{I}}_0, \emph{\textbf{I}}_1,
\emph{\textbf{I}}_2, \emph{\textbf{I}}_3)$. The radius vector in the
octonion space is $\mathbb{R} = \Sigma ( r_i \emph{\textbf{i}}_i +
R_i \emph{\textbf{I}}_i ) $ . The octonion velocity is $\mathbb{V} =
\Sigma ( v_i \emph{\textbf{i}}_i + V_i \emph{\textbf{I}}_i) $ .

When the electric charge is combined with the mass to become the
electron or the proton etc., we obtain the relation, $R_i
\emph{\textbf{I}}_i = r_i \emph{\textbf{i}}_i \circ
\emph{\textbf{I}}_0$, and $ V_i \emph{\textbf{I}}_i = v_i
\emph{\textbf{i}}_i \circ \emph{\textbf{I}}_0$. Meanwhile, the
gravitational intermediate boson and electromagnetic intermediate
boson may be combined together to become the photon etc. Here, the
symbol $\circ$ denotes the octonion multiplication.

The potential of gravitational field and electromagnetic field are
$\mathbb{A}_g = \Sigma ( a_i \emph{\textbf{i}}_i)$ and $\mathbb{A}_e
= \Sigma ( A_i \emph{\textbf{I}}_i)$ respectively. They are combined
together to become the potential in the octonion space.
\begin{eqnarray}
\mathbb{A} = \mathbb{A}_g + k_{eg} \mathbb{A}_e
\end{eqnarray}
where, $k_{eg}$ is a coefficient; $\mathbb{A}_e = \mathbb{A}_q \circ
\emph{\textbf{I}}_0$.

The strength $\mathbb{B}$ consists of the gravitational strength
$\mathbb{B}_g$ and electromagnetic strength $\mathbb{B}_e$ . The
selecting gauge equations are $b_0 = 0$ and $B_0 = 0$.
\begin{eqnarray}
\mathbb{B} = \lozenge \circ \mathbb{A} = \mathbb{B}_g + k_{eg}
\mathbb{B}_e
\end{eqnarray}

\begin{table}[t]
\caption{\label{tab:table1}The octonion multiplication table.}
\begin{ruledtabular}
\begin{tabular}{ccccccccc}
& 1  & $\emph{\textbf{i}}_1$ & $\emph{\textbf{i}}_2$ &
$\emph{\textbf{i}}_3$  & $\emph{\textbf{I}}_0$  &
$\emph{\textbf{I}}_1$
& $\emph{\textbf{I}}_2$  & $\emph{\textbf{I}}_3$  \\
\colrule 1 & $1$ & $\emph{\textbf{i}}_1$  & $\emph{\textbf{i}}_2$ &
$\emph{\textbf{i}}_3$  & $\emph{\textbf{I}}_0$  &
$\emph{\textbf{I}}_1$
& $\emph{\textbf{I}}_2$  & $\emph{\textbf{I}}_3$  \\
$\emph{\textbf{i}}_1$ & $\emph{\textbf{i}}_1$ & -1 &
$\emph{\textbf{i}}_3$  & $-\emph{\textbf{i}}_2$ &
$\emph{\textbf{I}}_1$
& $-\emph{\textbf{I}}_0$ & $-\emph{\textbf{I}}_3$ & $\emph{\textbf{I}}_2$  \\
$\emph{\textbf{i}}_2$ & $\emph{\textbf{i}}_2$ &
$-\emph{\textbf{i}}_3$ & -1 & $\emph{\textbf{i}}_1$  &
$\emph{\textbf{I}}_2$  & $\emph{\textbf{I}}_3$
& $-\emph{\textbf{I}}_0$ & $-\emph{\textbf{I}}_1$ \\
$\emph{\textbf{i}}_3$ & $\emph{\textbf{i}}_3$ &
$\emph{\textbf{i}}_2$ & $-\emph{\textbf{i}}_1$ & -1 &
$\emph{\textbf{I}}_3$  & $-\emph{\textbf{I}}_2$
& $\emph{\textbf{I}}_1$  & $-\emph{\textbf{I}}_0$ \\
\hline $\emph{\textbf{I}}_0$ & $\emph{\textbf{I}}_0$ &
$-\emph{\textbf{I}}_1$ & $-\emph{\textbf{I}}_2$ &
$-\emph{\textbf{I}}_3$ & -1 & $\emph{\textbf{i}}_1$
& $\emph{\textbf{i}}_2$  & $\emph{\textbf{i}}_3$  \\
$\emph{\textbf{I}}_1$ & $\emph{\textbf{I}}_1$ &
$\emph{\textbf{I}}_0$ & $-\emph{\textbf{I}}_3$ &
$\emph{\textbf{I}}_2$  & $-\emph{\textbf{i}}_1$
& -1 & $-\emph{\textbf{i}}_3$ & $\emph{\textbf{i}}_2$  \\
$\emph{\textbf{I}}_2$ & $\emph{\textbf{I}}_2$ &
$\emph{\textbf{I}}_3$ & $\emph{\textbf{I}}_0$  &
$-\emph{\textbf{I}}_1$ & $-\emph{\textbf{i}}_2$
& $\emph{\textbf{i}}_3$  & -1 & $-\emph{\textbf{i}}_1$ \\
$\emph{\textbf{I}}_3$ & $\emph{\textbf{I}}_3$ &
$-\emph{\textbf{I}}_2$ & $\emph{\textbf{I}}_1$  &
$\emph{\textbf{I}}_0$  & $-\emph{\textbf{i}}_3$ &
$-\emph{\textbf{i}}_2$ & $\emph{\textbf{i}}_1$  & -1 \\
\end{tabular}
\end{ruledtabular}
\end{table}

The gravitational strength $\mathbb{B}_g$ in Eq.(2) includes two
components, $\textbf{g} = ( g_{01} , g_{02} , g_{03} ) $ and
$\textbf{b} = ( g_{23} , g_{31} , g_{12} )$, while the
electromagnetic strength $\mathbb{B}_e$ involves two parts,
$\textbf{E} = ( B_{01} , B_{02} , B_{03} ) $ and $\textbf{B} = (
B_{23} , B_{31} , B_{12} )$.
\begin{eqnarray}
\textbf{E}/V_0 = && \emph{\textbf{I}}_1 ( \partial_0 A_1 +
\partial_1 A_0 ) + \emph{\textbf{I}}_2 ( \partial_0 A_2 + \partial_2
A_0 )
\nonumber\\
&& + \emph{\textbf{I}}_3 ( \partial_0 A_3 + \partial_3 A_0 )
\\
\textbf{B} = && \emph{\textbf{I}}_1 ( \partial_3 A_2 - \partial_2
A_3 ) + \emph{\textbf{I}}_2 ( \partial_1 A_3 - \partial_3 A_1 )
\nonumber\\
&& + \emph{\textbf{I}}_3 ( \partial_2 A_1 - \partial_1 A_2 )
\end{eqnarray}

The electric current density $\mathbb{S}_e = q \mathbb{V}_e$ is the
source of electromagnetic field in the octonion space. While the
linear momentum density is the source of gravitational field still.
And the source $\mathbb{S}$ was devised to consistently describe the
sources of electromagnetism and gravitation.
\begin{eqnarray}
\lozenge^* \circ \mathbb{B} = - \mu \mathbb{S} = - (\mu_g
\mathbb{S}_g + k_{eg} \mu_e \mathbb{S}_e)
\end{eqnarray}
where, $\mu$ is a coefficient; $\mu_e = \mu_q$; $k_{eg}^2 = \mu_g /
\mu_e$; $*$ denotes the conjugate of octonion.

From Eq.(31), we have,
\begin{eqnarray}
\lozenge^* \circ \mathbb{B}_g + k_{eg} \lozenge^* \circ \mathbb{B}_e
= - (\mu_g \mathbb{S}_g + k_{eg} \mu_e \mathbb{S}_e)
\end{eqnarray}
and the above equation can be decomposed as follows, according to
the basis vectors and multiplication.
\begin{eqnarray}
\lozenge^* \circ \mathbb{B}_g = - \mu_g \mathbb{S}_g
\\
\lozenge^* \circ \mathbb{B}_e = - \mu_e \mathbb{S}_e
\end{eqnarray}

In the octonion space, Eq.(33) is the same as Eq.(5) for the
gravitational field. And Eq.(34) is for the electromagnetic field.

The above description states that the electromagnetic field is
independent of the gravitational field absolutely, when the operator
$\lozenge$ is only dealt with the quaternion space but octonion
space. And that Maxwell equation of electromagnetic field remains
the same as that in classical electromagnetic theory in the octonion
space.

\subsection{Gauss's law}

In the electromagnetic field, the part $\textbf{S}_0$ of the source
$\mathbb{S}_e$ in Eq.(34) is rewritten as,
\begin{eqnarray}
\nabla^* \cdot \textbf{H} = - \mu_e \textbf{S}_0~.
\end{eqnarray}
where, $\textbf{S}_0 = S_0 \emph{\textbf{I}}_0$, $S_0 = q V_0$,
$\textbf{H} = \Sigma B_j \emph{\textbf{I}}_j $.

Further, the above is reduced to
\begin{eqnarray}
\nabla^* \cdot ( \textbf{E} / V_0 + \textbf{B} ) = - \mu_e q V_0
\emph{\textbf{I}}_0~.
\end{eqnarray}

Eqs.(28) and (30) yield the Gauss's law of magnetism,
\begin{eqnarray}
\nabla \cdot \textbf{B} = 0
\end{eqnarray}
and then, we have the Gauss's law as follows.
\begin{eqnarray}
\nabla^* \cdot \textbf{E} = - (q / \varepsilon_e)
\emph{\textbf{I}}_0
\end{eqnarray}
where, the coefficient $\varepsilon_e = 1/(\mu_e V_0^2)$.

The above states that the electromagnetic potential has an influence
on Gauss's law of electromagnetism. By far, we have two equations,
Eqs.(37) and (38), for the Maxwell equation, although the gauge
equation $B_0 = 0$ is different to that in classical electromagnetic
theory.

\begin{table}[t]
\caption{\label{tab:table1}The operator and multiplication of the
physical quantity in the octonion space.}
\begin{ruledtabular}
\begin{tabular}{ll}
definition                  &   meaning                                             \\
\hline
$\nabla \cdot \textbf{S}$   &  $-(\partial_1 S_1 + \partial_2 S_2 + \partial_3 S_3) \emph{\textbf{I}}_0 $  \\
$\nabla \times \textbf{S}$  &  $-\emph{\textbf{I}}_1 ( \partial_2
                                 S_3 - \partial_3 S_2 ) - \emph{\textbf{I}}_2 ( \partial_3 S_1
                                 - \partial_1 S_3 )$                                 \\
$$                          &  $ - \emph{\textbf{I}}_3 ( \partial_1 S_2
                                 - \partial_2 S_1 )$                                 \\
$\nabla \textbf{S}_0$       &  $\emph{\textbf{I}}_1 \partial_1 S_0
                                 + \emph{\textbf{I}}_2 \partial_2 S_0
                                 + \emph{\textbf{I}}_3 \partial_3 S_0  $             \\
$\partial_0 \textbf{S}$     &  $\emph{\textbf{I}}_1 \partial_0 S_1
                                 + \emph{\textbf{I}}_2 \partial_0 S_2
                                 + \emph{\textbf{I}}_3 \partial_0 S_3  $             \\
\end{tabular}
\end{ruledtabular}
\end{table}

\subsection{Ampere-Maxwell law}

In the octonion space, the vectorial part $\textbf{S}$ of
electromagnetic source $\mathbb{S}_e$ can be decomposed from
Eq.(34).
\begin{eqnarray}
\partial_0 \textbf{H} + \nabla^* \times \textbf{H} = - \mu_e \textbf{S}
\end{eqnarray}
where, $\textbf{S} = \Sigma (S_j \emph{\textbf{I}}_j) $; $S_j = q
V_j $ .

The above can be rewritten as follows.
\begin{eqnarray}
\partial_0 ( \textbf{E} / V_0 + \textbf{B} ) + \nabla^* \times ( \textbf{E} / V_0 + \textbf{B} )
= - \mu_e \textbf{S}
\end{eqnarray}

Eqs.(27), (29), and (30) yield the Faraday's law,
\begin{eqnarray}
\partial_0 \textbf{B} + \nabla^* \times \textbf{E} / V_0 = 0
\end{eqnarray}
and then, we obtain Ampere-Maxwell law in the electromagnetic field
as follows.
\begin{eqnarray}
\partial_0 \textbf{E} / V_0 + \nabla^* \times \textbf{B} = - \mu_e \textbf{S}
\end{eqnarray}

The above means that the electric charge in either steady state or
movement state can exert the electric force and magnetic force on
other charges. The strength $\textbf{B}$ and $\textbf{E}$ can be
induced each other in the electromagnetic field from Eq.(41). And
Eqs.(41) and (42) are combined with Eqs.(37) and (38) to become
Maxwell equation in the octonion space. With the relation, $
\mathbb{A}_e = \mathbb{A}_q \circ \emph{\textbf{I}}_0$, these
equations can be reduced to Maxwell equation in the quaternion
space.

\section{CONCLUSIONS}

The feature of gravitational field can be described with quaternion
spaces. Making use of the algebra of quaternions, we obtain the
Newton's law of gravitation etc. In the gravitational field, there
may exist the gravitational strength $\textbf{b}$, although the
strength part $\textbf{b}$ may be quite weak and difficult to
detect.

In the electromagnetic field, some characteristics can be
represented in the quaternion space too. By means of the algebra of
quaternions, we attain Maxwell equations etc. By comparison with
classical electromagnetic theory, we find the definition of gauge
equation is different, and the direction of displacement current in
Ampere-Maxwell equation is opposite.

With the octonion algebra, the gravitational field and
electromagnetic field can be described simultaneously. The
inferences in gravitational field or electromagnetic field can be
spread to the case for coexistence of electromagnetic field and
gravitational field. We achieve same conclusions as that in the
quaternion space, including the Maxwell equation etc.

It should be noted the study for the validity of Maxwell equation in
the electromagnetic field and gravitational field examined only some
simple cases in quaternion and octonion spaces. Despite its
preliminary characteristics, this study can clearly indicate that
the Newton's law of gravitation can be derived with quaternions. And
that Maxwell equation of electromagnetic field can be deduced with
the algebra of quaternions as well. In the octonion space, some
equations of the two fields can be drawn out simultaneously. For the
future studies, the research will concentrate on only some
predictions about the strong strength $\textbf{b}$ in gravitational
field as well as the direction of displacement current in
electromagnetic field.

\section*{Acknowledgments}
This project was supported partially by the National Natural Science
Foundation of China under grant number 60677039.

\end{document}